\documentclass[aps,prl,twocolumn,superscriptaddress,longbibliography]{revtex4-1}

\usepackage[version=3]{mhchem} 
\usepackage[T1]{fontenc}       
\usepackage{graphicx}
\usepackage{xcolor}
\usepackage[normalem]{ulem}
\usepackage{soul}

\begin{document}
\title{Packing polydisperse colloids into crystals: when charge-dispersity matters}

\author{Guillaume Bareigts}
\affiliation{ICB, CNRS UMR 6303, Univ. Bourgogne Franche-Comt\'e,
  21000 Dijon, France}
\author{Pree-Cha Kiatkirakajorn}
\affiliation{Max Planck Institute for Dynamics and Self-Organisation (MPIDS), G\"ottingen 37077, Germany}
\author{Joaquim Li}
\affiliation{LCMD, CNRS UMR 8231, ESPCI, 10 rue Vauquelin, 75231 Paris
  Cedex 05, France}
\author{Robert Botet}
\affiliation{Physique des Solides, CNRS UMR 8502, Univ Paris-Sud, F-91405 Orsay, France}
\author{Michael Sztucki}
\affiliation{ESRF-The European Synchrotron, CS40220, 38043 Grenoble Cedex 9, France}
\author{Bernard Cabane}
\affiliation{LCMD, CNRS UMR 8231, ESPCI, 10 rue Vauquelin, 75231 Paris Cedex 05, France}
\author{Lucas Goehring}
\email[]{lucas.goehring@ntu.ac.uk}
\affiliation{School of Science and Technology, Nottingham Trent University, Nottingham, NG11 8NS, UK}
\author{Christophe Labbez}
\email[]{Christophe.labbez@u-bourgogne.fr}
\affiliation{ICB, CNRS UMR 6303, Univ. Bourgogne Franche-Comt\'e, 21000 Dijon, France}

\date{\today}

\begin{abstract}

Monte-Carlo simulations, fully constrained by experimental parameters,
are found to agree well with a measured phase diagram of aqueous
dispersions of nanoparticles with a moderate size polydispersity over
a broad range of salt concentrations, $c_s$, and volume fractions,
$\phi$. Upon increasing $\phi$, the colloids freeze first into
coexisting compact solids then into a body centered cubic phase (bcc)
before they melt into a glass forming liquid. The surprising stability
of the bcc solid at high $\phi$ and $c_s$ is explained by  the
interaction (charge) polydispersity and vibrational entropy.

\end{abstract}

\maketitle

\begin{figure}
\includegraphics[width=85 mm]{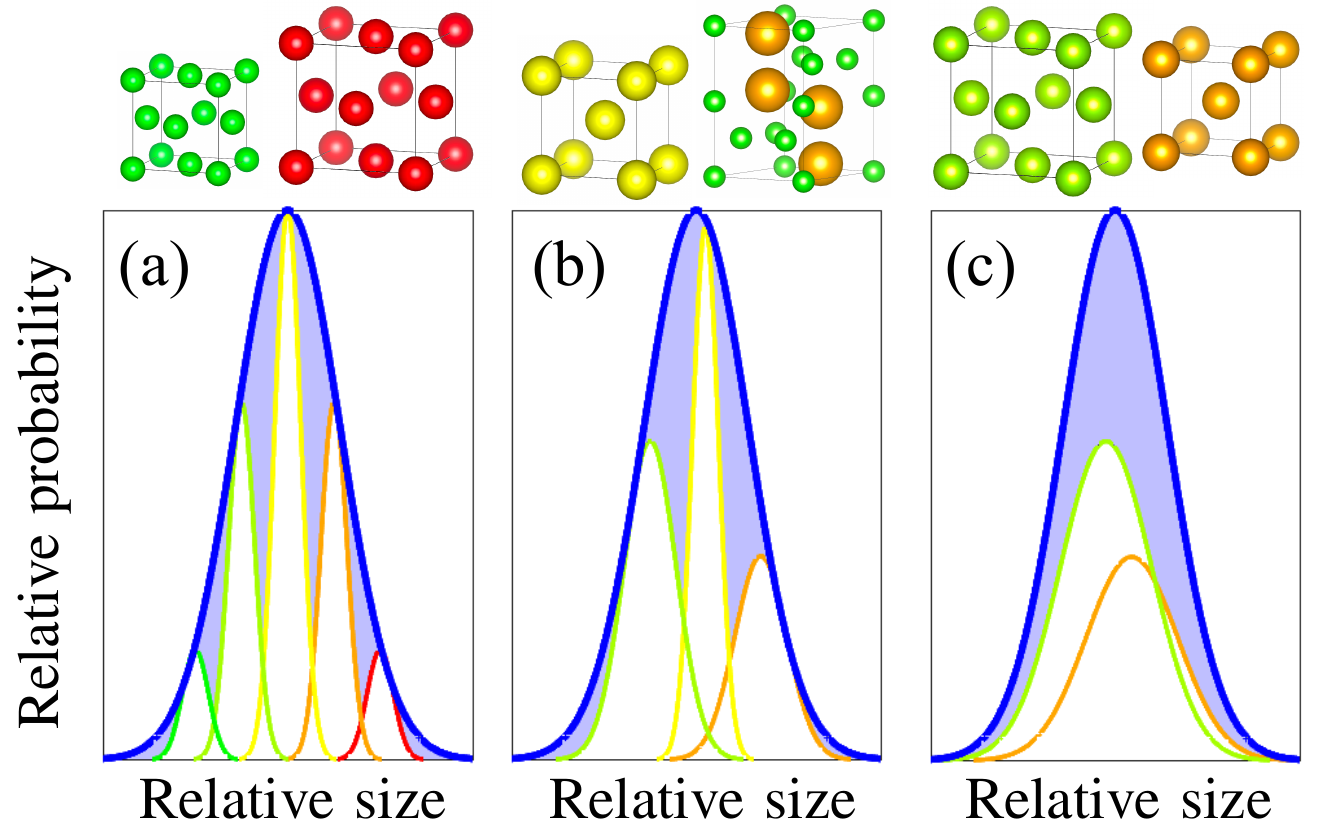}
\caption{Colloidal crystallisation in a polydisperse system can lead
  to: (a) A set of distinct crystals of the same structure (e.g. fcc)
  and narrow monomodal size distributions, which together span the
  available range of particle sizes \cite{Fasolo:03}; (b) More
  complex phases such as AB$_2$ \cite{Cabane:16} or AB$_{13}$
  \cite{Bommineni:19} structures, which utilise a bimodal subset of
  particles. These may coexist with simpler phases (e.g. as above, bcc
  \cite{Cabane:16}); (c) The appearance of crystals of different
  structures (e.g. bcc, fcc, hcp) and monomodal size distributions, as
  reported in this paper.  In all sketches the shaded area shows the parent particle size distribution while the various open curves describe the particles found in any specific crystal structure and site.}
\label{fig:routes}
\end{figure}

How do polydisperse particles pack and order? This basic question concerns diverse systems, including granular beads,
micro-emulsions, micro-gels, macromolecules and solid nanoparticles
and is, thus, largely debated.  For a fluid of hard-sphere (HS)
particles, Pusey et al.~\cite{Pusey:87,Pusey:91} proposed a critical
value of polydispersity ($\delta$), above which particles would not
crystallize.  This concept of a terminal polydispersity was first
based on experimental observations, and later supported also by
numerical simulations \cite{Auer:01,Pusey:09}. However, using
simulations of HS systems, Kofke et al.~\cite{Kofke:99} found that the
concept of a terminal polydispersity should only apply to a solid
phase, rather than the entire system of particles. More precisely,
that a stable crystalline phase whose constituent components exceeded a
polydispersity of 5.7$\%$ could not be formed from a fluid
phase. Questioning the ultimate fate of an amorphous solid of high
$\delta$, they proposed that fractionation should enable an HS fluid
of arbitrary polydispersity to precipitate in a fcc solid phase in
coexistence with a fluid phase. Sollich et
al.~\cite{Fasolo:03,Sollich:10} further theorized that, when
compressed, a relatively polydisperse HS system should crystallize
into a myriad of coexisting fcc crystalline phases each having a
distinct size distribution and a narrower $\delta$ than the mother
distribution, as in Fig.~\ref{fig:routes}(a).

Our recent experiments \cite{Cabane:16} on dispersions of charged hard
spheres (CS) with a broad and continuous size polydispersity ($\delta
= 14\%$) empirically demonstrated the case of the fractionation of a
colloidal fluid into multiple coexisting phases. Interestingly, this
crystallization turns out to be more complex than that theorized by
Sollich et al. for HS. Indeed, as in Fig.~\ref{fig:routes}(b), the CS
were observed to coexist in a fluid phase, a bcc lattice and a Laves
MgZn$_2$ superlattice. The latter had been previously known only from
\textit{binary} distributions of particles
\cite{Shevchenko:06,Hynninen:07,Schaertl:18}. Matching lattice
simulations can also reproduce the experimental findings, including
the Laves phase \cite{Botet:16,Cabane:16}. Very recent simulations
\cite{Coslovich:18,Lindquist:18,Bommineni:19} with polydisperse HS of
$\delta > 6\%$ show a similar, or even greater, level of complexity
and thus indicate that our findings with CS are representative of a
more general rule: polydispersity enables complex crystal
formation. In particular, Frank-Kasper phases, as well as various
Laves AB$_{2}$ and AB$_{13}$ phases were found in simulations of HS of
$\delta$ from 6\% to 24\% and at high packing fractions
($\phi$). These results are also in line with the earlier simulations
of Fernandez el al.~\cite{Fernandez:07}, of neutral soft spheres, even
though the exact natures of the crystal phases obtained there were not
identified. On the other hand, the coexistence of
multiple crystal phases of the same symmetry, but different lattice constants, has only been observed in systems of plate-like particles~\cite{Byelov:10}.

Here, we demonstrate that even with a moderate size polydispersity CS
systems can show a complex phase behavior. This is achieved on a
similar CS system to that in~\cite{Cabane:16} but with a moderate size
dispersity (9\%). The magnitude and polydispersity of the charge, and
thus of the interaction polydispersity, are tuned with
the salt concentration, $c_s$ and pH of the bulk solution (see
Supplemental Material \cite{NoteSM}). Using x-ray scattering methods the
$c_s$ -- $\phi$ phase diagram is constructed. We observe that on gradually increasing the osmotic compression the
CS fluid crystallises and fractionates into coexisting phases of different
structures, i.e. bcc, fcc and hcp, as in Fig.~\ref{fig:routes}(c). Unexpectedly, the stability region
of the bcc crystals covers a large area of the
phase diagram, considerably more than in the monodisperse case.   The first appearance of the bcc phase is always at a higher $\phi$ than that of fcc crystals, at the same $c_s$ (i.e. opposite to their order of occurrence in monodisperse CS \cite{Hynninen:03}). Upon further compression, the system
becomes a glass forming liquid. To help explain
these results, we use Monte Carlo (MC)
simulations of our multi-component model (MCM) for
charge regulating polydisperse colloids parametrized with independent
experimental data~\cite{Bareigts:18}. Allowing for only a slight adjustment of $\delta$, the
simulations almost perfectly reproduce the experimental phase diagram.

\begin{figure}
\includegraphics[width=85 mm]{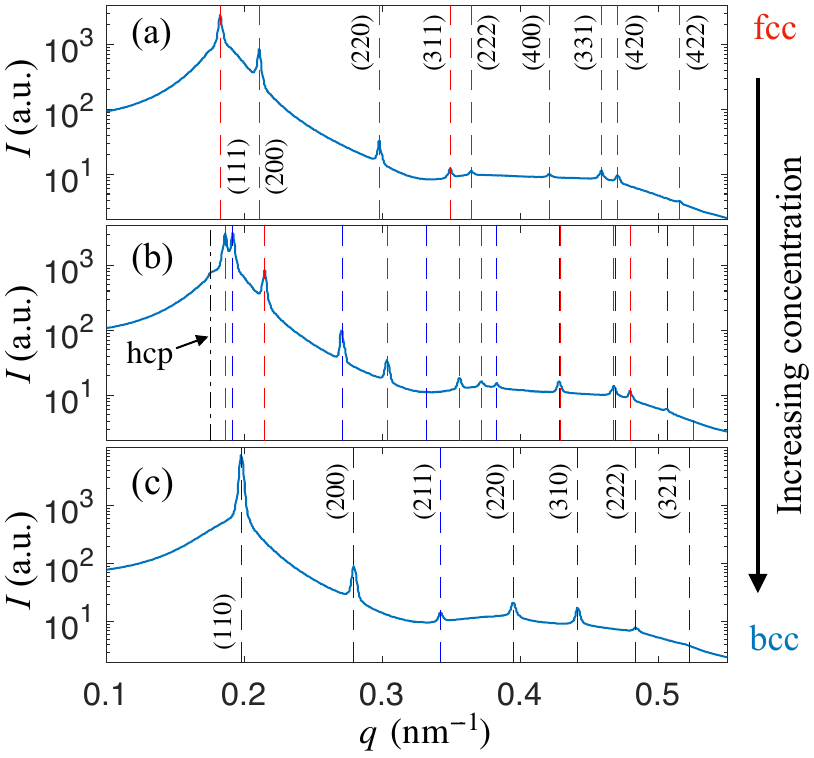}
  \caption{As the dispersion is concentrated, colloidal crystals
    appear.  The scattering intensities, $I$, of the spectra shown here, for $c_s = 5$ mM,
    demonstrate the typical sequence of (a) fcc ($\phi = 19\%$), (b) a
    mixture of fcc and bcc ($\phi = 20\%$) and (c) bcc ($\phi = 21\%$)
    crystals, as $\phi$ increases.  A broad liquid peak is present in all spectra, and the most prominent crystal peaks are typically at least twice as intense as this liquid background.  Additionally, a much weaker
    peak is often visible at lower $q$, consistent with an hcp structure of
    the same particle density, or stacking faults in an fcc
    lattice.   
  \label{fig:methods}}
\end{figure}

For the experiments, we used industrially produced, nanometric and highly charged silica
particles, dispersed in water (Ludox TM50, Sigma-Aldrich).  These were
cleaned and concentrated as detailed elsewhere
\cite{Jonsson:11a,Li:15,Cabane:16,Goehring:17}.  Briefly, dispersions
were filtered and dialysed against aqueous NaCl solutions of various
concentrations (from 0.5 to 50 mM) at pH $9\pm0.5$ (by addition of NaOH).  Next, they were
slowly concentrated via the osmotic stress method, by the addition of
polyethylene glycol (mw~35000, Sigma-Aldrich) outside the dialysis
sack.  Samples were then taken and sealed in quartz capillary tubes,
on which small-angle x-ray scattering (SAXS) experiments were
performed at the ESRF, beamline ID02 \cite{Narayanan:18}.  The particle size distribution
was measured in the dilute limit (see the Supplemental Material \cite{NoteSM}) to
have a mean size of $\overline{R}=13.75\pm$1 nm and a polydispersity
of $\delta = 9\pm$1\%, consistent with prior
observations~\cite{Goertz:09}.  Over a range of concentrations
the scattering spectra showed sharp peaks characteristic of fcc and
bcc crystal phases, as shown in Fig.~\ref{fig:methods}.  A weak peak
representing a minority hcp phase (or evidence of stacking faults
\cite{Shabalin:16}) was frequently seen alongside either crystal
phase.  Additional characterisation of the liquid and glass phases is given in the Supplemental Material \cite{NoteSM}.

\begin{figure}[!tbp]
\includegraphics[width=85 mm]{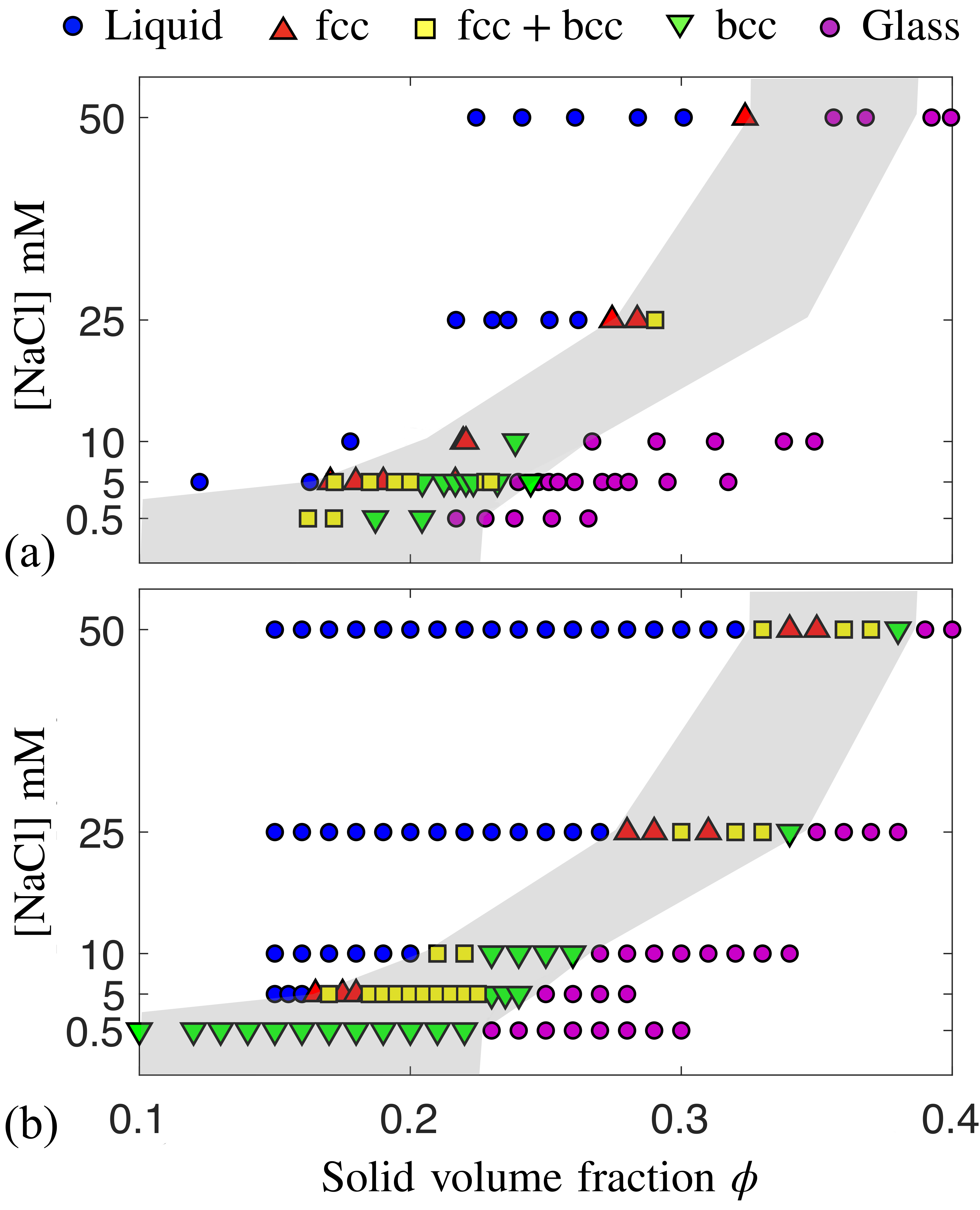}
  \caption{Phase diagram in the $c_s-\phi$ plane of the TM50 silica
    dispersion at pH 9 as obtained from (a) SAXS analysis of dialysed samples and (b) MC simulations of the MCM. For clarity, the hcp phase is not represented. The shaded areas of (a,b) delimit the region
    where crystals are found in the simulations, and demonstrate the good correspondence with experiments.
  } 
\label{fig:phase}
\end{figure}

The experimental phase diagram in the $c_s$ -- $\phi$ plane
is given in Fig.~\ref{fig:phase}(a), and represents the phases that have nucleated and are experimentally stable over days-to-weeks. Whatever the background salinity, a
fluid region is observed for low $\phi$ followed by a region with
crystal formation at intermediate $\phi$, which ends in a re-entrant amorphous
phase at high $\phi$. The latter behaves macroscopically as a
solid (i.e. retains its shape as a soft gel or paste).   As $c_s$ is increased, the first appearance of crystals shifts to higher $\phi$, in response to the screening of the electrostatic interactions. The
same is true for the re-entrant melting transition.  Both observations are consistent with
phase diagrams of other experimental CS systems although at much lower
$c_s$, (e.g. \cite{Monovoukas:89,Sirota:89}). The predominant ordered phases
appearing are bcc and fcc crystals, also known from
monodisperse CS systems (although the hcp phase is not typically seen
there, other than in shear-ordered samples \cite{Versmold:95}).
However, the stability region of the bcc phase is observed at higher
$\phi$ than the fcc phase, for all screening lengths studied (i.e. all
$c_s$). This was also the case where we made a more continuous probe
across $\phi$ as assessed by interdiffusion experiments (see the
Supplemental Material \cite{NoteSM} for
details). This phase behavior contrasts strongly with that of
monodisperse particles, where a \textit{bcc-fcc} transition with
increasing $\phi$ (rather than the \textit{fcc-bcc} transition seen
here) is invariably observed
\cite{Monovoukas:89,Sirota:89}. Essentially, this demonstrates that
even a moderate polydispersity can have a complex influence on crystal
stability, and modify the relative stability of various phases.


Although predicted to occur for soft
colloids \cite{Gottwald:04,Pamies:09} an inversion of the stability
regions of the bcc and fcc phases has rarely been observed. To our
knowledge, it has only been reported for soft spheres \cite{Mohanty:08}. The possibility that polydispersity could help stabilise the bcc phase in CS systems was conjectured by some of us, based on an energetic argument which shows that
the bcc structure is more tolerant to interaction polydispersity than
the fcc one \cite{Botet:16}. This argument was made via \textit{lattice} MC
simulations in the Gibbs ensemble on a system with a presupposed bcc/fcc
coexistence of CS with $\delta$ = 15\%.  The particles were found to be divided up between a narrow monomodal distribution (fcc) and a bimodal one (bcc), similar to the situation given in Fig~\ref{fig:routes}(b).  However, our reanalysis of this model shows that it also predicts the formation of a CsCl structure on the sites of the bcc phase (i.e. alternating larger and smaller particles), which is not compatible with our experimental findings (as additional scattering peaks would be present in this case). 

\begin{figure}[!htbp]
\includegraphics[width=85 mm]{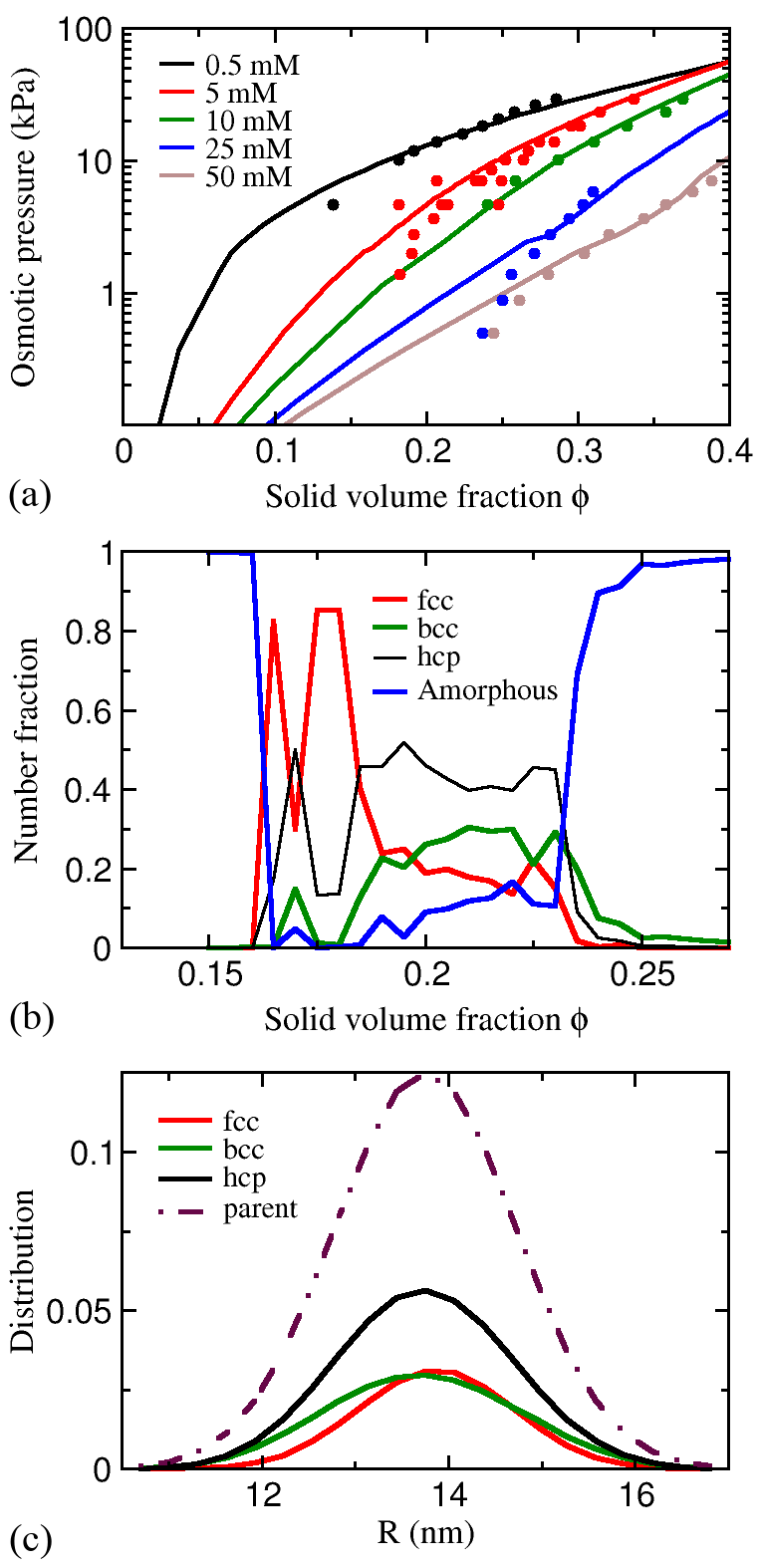}
\caption{ Simulation results (at pH 9) show: (a) agreement of the
  simulated and measured equation of state (EoS) of the TM50
    silica dispersion at various ionic strengths; (b)
    the predicted variation of the phase composition with $\phi$ at $c_s=$
    5 mM; (c) and particle size distributions of the various crystalline phases in comparison with the
    parent size distribution (dashed curve) for the model silica dispersion at
    $\phi$ = 20.5\%, $c_s=$ 5 mM.}
	\label{fig:simdet}
\end{figure}

Here we employed, instead, MC simulations for \textit{continuous} systems at
set density (NVT) or pressure (NPT) which do not require any prior
information on the phases at equilibrium. They
were performed at the experimental $c_s$ and pH
conditions in the framework of the MCM detailed in Ref.~\cite{Bareigts:18},
which includes the charge regulation of the silica particles
through the pH dependent ionization of their surface active groups,
$\ce{Si-OH \ce{<=>} Si-O^{-} + H^{+}}$. A truncated
and discretized Gaussian size distribution with the same
$\overline{R}$ as measured was used, but with a somewhat lower polydispersity of $\delta = 7\%$ (rather than 9\%). Simple particle translations combined with swap moves \cite{Grigera:01} allow for efficiently sampling the phase space up to high
$\phi$ \cite{Brito:18}. Simulations were run with $N$ = 19 991 particles
in a cubic box with periodic boundary conditions. Up to several tens of million
of MC cycles (each consisting of $N$ MC moves) for equilibration were used; production
runs lasted for 10$^5$ MC cycles. The local bond order parameters were used
to analyze the obtained structures \cite{Lechner:08}, and further
details of the analysis and simulation are given in the Supplemental
Material \cite{NoteSM}.

As shown in Fig.~\ref{fig:phase}(b), a very good agreement is
achieved between the experimental and simulated (MCM) phase diagrams. The
same is true for the equation of state (EoS) of the TM50 silica dispersion
in the all range of $c_s$ and $\phi$ studied as seen in
Fig~\ref{fig:simdet}(a) (experimental data from
Refs. \cite{Goehring:17,Kiatkirakajorn:18}). Not only is the inversion
of the stability regions of the bcc and fcc phases well predicted, but
also the position of the freezing transition matches with the experiments,
although an exact phase diagram would require a free energy
calculation not developed here (see the Supplemental Material for
discussion). In line with the experimental observations
\cite{Goehring:17}, a re-entrant amorphous phase at high $\phi$ is
found, in which colloids present very weak diffusion.

The phase composition of the system upon compression at $c_s=$ 5 mM
and an example of the size distributions at the coexistence of the
hcp/bcc/fcc phases are shown in Fig~\ref{fig:simdet}(b,c). The
freezing transition is found to be first-order, and is characterized by both a
discontinuity in the EoS and an abrupt change in the liquid/fcc phase
composition at $\phi \approx 16\%$, as in Fig~\ref{fig:simdet}(a). The fcc-bcc phase transition is, on the other hand, found to be much more
progressive. Simulation snapshots at the bcc/fcc/hcp phase coexistence
show, instead, textures characteristic of a micro-phase separation
(for further detail, see the Supplemental Material \cite{NoteSM}). 


The fcc-bcc phase transition is also characterized by a
small size fractionation, as in Fig~\ref{fig:simdet}(c), which tends to increase with $\phi$ (see the Supplemental Material \cite{NoteSM}). The bcc phase is found to be more tolerant to polydispersity, while incorporating a
larger number of small particles than the fcc structure.
The particle distribution of the bcc phase thus presents a
larger $\delta$ and smaller $R$ than of the fcc phase.
One consequence of this is that there will be only a small difference in the calculated particle number
densities between both crystalline phases and the bulk, less than
4\%, with a slight tendency of the fcc phase to be the densest.
This is consistent with our experimental observations,
although the difference in phase densities falls within the
uncertainty of the measurements (see the Supplemental Material \cite{NoteSM}).

These results are in line with our energetic argument
mentioned earlier \cite{Botet:16}. In other words, the interaction
polydispersity favors the formation of bcc crystals with a larger
particle distribution (or charge distribution), thus being
more tolerant to polydispersity, as compared to fcc crystals. As
$\phi$ is progressively increased, the fcc phase, compared to the bcc
phase, becomes less and less tolerant to the charge
polydispersity. Note that the latter is not constant but increases
with $\phi$, see \cite{Bareigts:18}. Consequently, the fcc ordered
phase progressively disappears in favor of the bcc and fluid
phases. Conversely, in the absence of interaction polydispersity the
system can, to a good approximation, be reduced to that of point
Yukawa particles \cite{Hynninen:03}. In such a case, the charge of
colloid $i$ satisfies the equality $Z_i^{\ast}\exp(-\kappa^{\ast}
R_i)/(1+\kappa^{\ast}R_i)=\text{C}\neq 0$ $\forall$ $R_i$, where $C$ is a constant (see the Supplemental Material \cite{NoteSM}). The inversion of
the stability regions of the bcc and fcc phases is then lost
\cite{Hamaguchi:97}. In this case also, the stability region of the
bcc phase is restricted to the very diluted $c_s-\phi$ domain
only. In absence of charge (i.e. $C = 0$), the bcc phase simply
disappears, see e.g. the recent work of Bommineni on polydisperse HS
systems \cite{Bommineni:19}. All this further illustrates the
importance of charge dispersity in the inversion of the stability
regions of the bcc and fcc phases.

Obviously, the phase behavior observed in our
experiments and simulations is not only a consequence
of the system's internal energy, but the result of the balance
between energy and entropy. In an attempt to
elucidate the entropic contributions in the stabilization of the
bcc phase we further performed lattice simulations in
the Gibbs ensemble, as in Ref. \cite{Botet:16}, with the
MCM of the TM50 silica dispersion. As in the continuous simulations, a
small size fractionation is obtained. However, a CsCl superlattice
structure, instead of a bcc phase, is found (see the Supplemental
Material \cite{NoteSM}). Recognizing
that lattice simulations only account for the mixing contribution to the entropy,
one can deduce from this qualitative difference that the
bcc phase observed in our experiments (and continuous simulations) is
most probably stabilized by vibrational entropy (the missing thermodynamic ingredient
in lattice simulations). A large size
fractionation in distinct phases is, on the other hand,
prevented by the mixing entropy at this relatively small $\delta$ and
range of $\phi$. When the size polydispersity is increased (see the
Supplemental Material \cite{NoteSM}), the mixing entropy takes over,
and a MgZn$_2$
Laves phase in coexistence with a bcc phase is predicted to occur
in good agreement with our previous experimental findings~\cite{Cabane:16}\nocite{Botet:12a,Li:12,Boon:15,Bolt:57,Labbez:09,Dove:05,Frenkel:02,Wilding:10,Verlet:68,Hansen:69,Leocmach:13}.

Not discussed so far is the striking agreement obtained between the simulations and experiments on
the position of the re-entrant melting line. At a first sight, this
would suggest that the amorphous phase is
stable. Preliminary results obtained well inside the amorphous region
with more advanced simulation techniques show, however, that it can
crystallize. A close look at the EoS also shows a sudden increase in
the osmotic pressure. These results, which will be developed elsewhere, strongly
suggest that it is a glass forming liquid. Still, we were unable to
come with a reasonable explanation for the troubling coincidence
between our simulation and experimental results on the
(non-thermodynamic) re-entrant melting transition.

To conclude, using a combined and detailed theoretical and experimental study
of charged nano-colloids with a moderate polydispersity, we provide
evidence that the packing of polydisperse particles into crystals is much more diverse
than initially thought, even for relatively small polydispersities. In
particular, the system is found to separate into coexisting solid
phases with a limited size fractionation. Under compression, the
system first solidifies in compact lattice structures, fcc/hcp. Upon
further compression, the fcc phase dissolves progressively into a less
compact bcc structure, which proves to be more tolerant to the interaction (charge)
polydispersity. Our simulations strongly suggest that the limited size
fractionation and the stabilization of the bcc phase are due to the
mixing and vibrational entropies, respectively. Compressed even
further, the colloidal crystals melt into an amorphous phase, most probably a glass forming liquid. The
astonishingly good agreement obtained between our experimental results
and simulated predictions further gives a strong support to the
simulation methods employed and the parameter-free force field
developed. We anticipate that these tools should help in the finding of
new colloidal crystal phases and in providing a better understanding of
colloidal glasses in CS systems. Still, the exact phase boundaries and
equilibrium phase behavior of polydisperse CS, in particular at high
densities, remain open questions which will require the development of
advanced simulation techniques to be tackled.

\begin{acknowledgments}
We acknowledge the European Synchrotron Radiation Facility for provision of synchrotron radiation facilities and assistance in using beamline ID02.
G.B. and C.L. thanks financial support from the Region Bourgogne
Franche-Comt\'e and CNRS, as well as computational support from CRI, Universit\'e de Bourgogne. P.-C.K. thanks the Thai DPST and the Royal Government of Thailand for funding.  G.B. and P.-C.K. contributed equally to this work.
\end{acknowledgments}




\bibliography{article}

\begin{thebibliography}{48}%
\makeatletter
\providecommand \@ifxundefined [1]{%
 \@ifx{#1\undefined}
}%
\providecommand \@ifnum [1]{%
 \ifnum #1\expandafter \@firstoftwo
 \else \expandafter \@secondoftwo
 \fi
}%
\providecommand \@ifx [1]{%
 \ifx #1\expandafter \@firstoftwo
 \else \expandafter \@secondoftwo
 \fi
}%
\providecommand \natexlab [1]{#1}%
\providecommand \enquote  [1]{``#1''}%
\providecommand \bibnamefont  [1]{#1}%
\providecommand \bibfnamefont [1]{#1}%
\providecommand \citenamefont [1]{#1}%
\providecommand \href@noop [0]{\@secondoftwo}%
\providecommand \href [0]{\begingroup \@sanitize@url \@href}%
\providecommand \@href[1]{\@@startlink{#1}\@@href}%
\providecommand \@@href[1]{\endgroup#1\@@endlink}%
\providecommand \@sanitize@url [0]{\catcode `\\12\catcode `\$12\catcode
  `\&12\catcode `\#12\catcode `\^12\catcode `\_12\catcode `\%12\relax}%
\providecommand \@@startlink[1]{}%
\providecommand \@@endlink[0]{}%
\providecommand \url  [0]{\begingroup\@sanitize@url \@url }%
\providecommand \@url [1]{\endgroup\@href {#1}{\urlprefix }}%
\providecommand \urlprefix  [0]{URL }%
\providecommand \Eprint [0]{\href }%
\providecommand \doibase [0]{http://dx.doi.org/}%
\providecommand \selectlanguage [0]{\@gobble}%
\providecommand \bibinfo  [0]{\@secondoftwo}%
\providecommand \bibfield  [0]{\@secondoftwo}%
\providecommand \translation [1]{[#1]}%
\providecommand \BibitemOpen [0]{}%
\providecommand \bibitemStop [0]{}%
\providecommand \bibitemNoStop [0]{.\EOS\space}%
\providecommand \EOS [0]{\spacefactor3000\relax}%
\providecommand \BibitemShut  [1]{\csname bibitem#1\endcsname}%
\let\auto@bib@innerbib\@empty
\bibitem [{\citenamefont {Pusey}(1987)}]{Pusey:87}%
  \BibitemOpen
  \bibfield  {author} {\bibinfo {author} {\bibfnamefont {P.~N.}\ \bibnamefont
  {Pusey}},\ }\bibfield  {title} {\enquote {\bibinfo {title} {The effect of
  polydispersity on the crystallization of hard spherical colloids},}\ }\href
  {\doibase 10.1051/jphys:01987004805070900} {\bibfield  {journal} {\bibinfo
  {journal} {J. Phys. France}\ }\textbf {\bibinfo {volume} {48}},\ \bibinfo
  {pages} {709--712} (\bibinfo {year} {1987})}\BibitemShut {NoStop}%
\bibitem [{\citenamefont {Pusey}(1991)}]{Pusey:91}%
  \BibitemOpen
  \bibfield  {author} {\bibinfo {author} {\bibfnamefont {P.~N.}\ \bibnamefont
  {Pusey}},\ }\bibfield  {title} {\enquote {\bibinfo {title} {Colloidal
  suspensions},}\ }in\ \href@noop {} {\emph {\bibinfo {booktitle} {Les
  {{Houches Session L1}}: {{Liquids}}, {{Freezing}}, and the {{Glass
  Transition}}}}}\ (\bibinfo  {publisher} {{edited by J. P. Hansen, D.
  Levesque, and J. Zinn-Justin (North-Holland, 1991)}},\ \bibinfo {year}
  {1991})\ pp.\ \bibinfo {pages} {765--942}\BibitemShut {NoStop}%
\bibitem [{\citenamefont {Auer}\ and\ \citenamefont {Frenkel}(2001)}]{Auer:01}%
  \BibitemOpen
  \bibfield  {author} {\bibinfo {author} {\bibfnamefont {Stefan}\ \bibnamefont
  {Auer}}\ and\ \bibinfo {author} {\bibfnamefont {Daan}\ \bibnamefont
  {Frenkel}},\ }\bibfield  {title} {\enquote {\bibinfo {title} {Suppression of
  crystal nucleation in polydisperse colloids due to increase of the surface
  free energy},}\ }\href@noop {} {\bibfield  {journal} {\bibinfo  {journal}
  {Nature}\ }\textbf {\bibinfo {volume} {413}},\ \bibinfo {pages} {711--713}
  (\bibinfo {year} {2001})}\BibitemShut {NoStop}%
\bibitem [{\citenamefont {Pusey}\ \emph {et~al.}(2009)\citenamefont {Pusey},
  \citenamefont {Zaccarelli}, \citenamefont {Valeriani}, \citenamefont {Sanz},
  \citenamefont {Poon},\ and\ \citenamefont {Cates}}]{Pusey:09}%
  \BibitemOpen
  \bibfield  {author} {\bibinfo {author} {\bibfnamefont {P.~N.}\ \bibnamefont
  {Pusey}}, \bibinfo {author} {\bibfnamefont {E.}~\bibnamefont {Zaccarelli}},
  \bibinfo {author} {\bibfnamefont {C.}~\bibnamefont {Valeriani}}, \bibinfo
  {author} {\bibfnamefont {E.}~\bibnamefont {Sanz}}, \bibinfo {author}
  {\bibfnamefont {Wilson C.~K.}\ \bibnamefont {Poon}}, \ and\ \bibinfo {author}
  {\bibfnamefont {Michael~E.}\ \bibnamefont {Cates}},\ }\bibfield  {title}
  {\enquote {\bibinfo {title} {Hard spheres: Crystallization and glass
  formation},}\ }\href {\doibase 10.1098/rsta.2009.0181} {\bibfield  {journal}
  {\bibinfo  {journal} {Philos. Trans. R. Soc. Lond. Math. Phys. Eng. Sci.}\
  }\textbf {\bibinfo {volume} {367}},\ \bibinfo {pages} {4993--5011} (\bibinfo
  {year} {2009})}\BibitemShut {NoStop}%
\bibitem [{\citenamefont {Kofke}\ and\ \citenamefont
  {Bolhuis}(1999)}]{Kofke:99}%
  \BibitemOpen
  \bibfield  {author} {\bibinfo {author} {\bibfnamefont {David~A.}\
  \bibnamefont {Kofke}}\ and\ \bibinfo {author} {\bibfnamefont {Peter~G.}\
  \bibnamefont {Bolhuis}},\ }\bibfield  {title} {\enquote {\bibinfo {title}
  {Freezing of polydisperse hard spheres},}\ }\href {\doibase
  10.1103/PhysRevE.59.618} {\bibfield  {journal} {\bibinfo  {journal} {Phys Rev
  E}\ }\textbf {\bibinfo {volume} {59}},\ \bibinfo {pages} {618--622} (\bibinfo
  {year} {1999})}\BibitemShut {NoStop}%
\bibitem [{\citenamefont {Fasolo}\ and\ \citenamefont
  {Sollich}(2003)}]{Fasolo:03}%
  \BibitemOpen
  \bibfield  {author} {\bibinfo {author} {\bibfnamefont {Moreno}\ \bibnamefont
  {Fasolo}}\ and\ \bibinfo {author} {\bibfnamefont {Peter}\ \bibnamefont
  {Sollich}},\ }\bibfield  {title} {\enquote {\bibinfo {title} {Equilibrium
  {{Phase Behavior}} of {{Polydisperse Hard Spheres}}},}\ }\href {\doibase
  10.1103/PhysRevLett.91.068301} {\bibfield  {journal} {\bibinfo  {journal}
  {Phys Rev Lett}\ }\textbf {\bibinfo {volume} {91}},\ \bibinfo {pages}
  {068301} (\bibinfo {year} {2003})}\BibitemShut {NoStop}%
\bibitem [{\citenamefont {Sollich}\ and\ \citenamefont
  {Wilding}(2010)}]{Sollich:10}%
  \BibitemOpen
  \bibfield  {author} {\bibinfo {author} {\bibfnamefont {Peter}\ \bibnamefont
  {Sollich}}\ and\ \bibinfo {author} {\bibfnamefont {Nigel~B.}\ \bibnamefont
  {Wilding}},\ }\bibfield  {title} {\enquote {\bibinfo {title} {Crystalline
  {{Phases}} of {{Polydisperse Spheres}}},}\ }\href {\doibase
  10.1103/PhysRevLett.104.118302} {\bibfield  {journal} {\bibinfo  {journal}
  {Phys. Rev. Lett.}\ }\textbf {\bibinfo {volume} {104}},\ \bibinfo {pages}
  {118302} (\bibinfo {year} {2010})}\BibitemShut {NoStop}%
\bibitem [{\citenamefont {Cabane}\ \emph {et~al.}(2016)\citenamefont {Cabane},
  \citenamefont {Li}, \citenamefont {Artzner}, \citenamefont {Botet},
  \citenamefont {Labbez}, \citenamefont {Bareigts}, \citenamefont {Sztucki},\
  and\ \citenamefont {Goehring}}]{Cabane:16}%
  \BibitemOpen
  \bibfield  {author} {\bibinfo {author} {\bibfnamefont {Bernard}\ \bibnamefont
  {Cabane}}, \bibinfo {author} {\bibfnamefont {Joaquim}\ \bibnamefont {Li}},
  \bibinfo {author} {\bibfnamefont {Franck}\ \bibnamefont {Artzner}}, \bibinfo
  {author} {\bibfnamefont {Robert}\ \bibnamefont {Botet}}, \bibinfo {author}
  {\bibfnamefont {Christophe}\ \bibnamefont {Labbez}}, \bibinfo {author}
  {\bibfnamefont {Guillaume}\ \bibnamefont {Bareigts}}, \bibinfo {author}
  {\bibfnamefont {Michael}\ \bibnamefont {Sztucki}}, \ and\ \bibinfo {author}
  {\bibfnamefont {Lucas}\ \bibnamefont {Goehring}},\ }\bibfield  {title}
  {\enquote {\bibinfo {title} {Hiding in {{Plain View}}: {{Colloidal
  Self}}-{{Assembly}} from {{Polydisperse Populations}}},}\ }\href {\doibase
  10.1103/PhysRevLett.116.208001} {\bibfield  {journal} {\bibinfo  {journal}
  {Phys. Rev. Lett.}\ }\textbf {\bibinfo {volume} {116}},\ \bibinfo {pages}
  {208001} (\bibinfo {year} {2016})}\BibitemShut {NoStop}%
\bibitem [{\citenamefont {Bommineni}\ \emph {et~al.}(2019)\citenamefont
  {Bommineni}, \citenamefont {{Varela-Rosales}}, \citenamefont {Klement},\ and\
  \citenamefont {Engel}}]{Bommineni:19}%
  \BibitemOpen
  \bibfield  {author} {\bibinfo {author} {\bibfnamefont {Praveen~K.}\
  \bibnamefont {Bommineni}}, \bibinfo {author} {\bibfnamefont {Nydia~Roxana}\
  \bibnamefont {{Varela-Rosales}}}, \bibinfo {author} {\bibfnamefont {Marco}\
  \bibnamefont {Klement}}, \ and\ \bibinfo {author} {\bibfnamefont {Michael}\
  \bibnamefont {Engel}},\ }\bibfield  {title} {\enquote {\bibinfo {title}
  {Complex {{Crystals}} from {{Size}}-{{Disperse Spheres}}},}\ }\href {\doibase
  10.1103/PhysRevLett.122.128005} {\bibfield  {journal} {\bibinfo  {journal}
  {Phys. Rev. Lett.}\ }\textbf {\bibinfo {volume} {122}},\ \bibinfo {pages}
  {128005} (\bibinfo {year} {2019})}\BibitemShut {NoStop}%
\bibitem [{\citenamefont {Shevchenko}\ \emph {et~al.}(2006)\citenamefont
  {Shevchenko}, \citenamefont {Talapin}, \citenamefont {Kotov}, \citenamefont
  {O'Brien},\ and\ \citenamefont {Murray}}]{Shevchenko:06}%
  \BibitemOpen
  \bibfield  {author} {\bibinfo {author} {\bibfnamefont {Elena~V.}\
  \bibnamefont {Shevchenko}}, \bibinfo {author} {\bibfnamefont {Dmitri~V.}\
  \bibnamefont {Talapin}}, \bibinfo {author} {\bibfnamefont {Nicholas~A.}\
  \bibnamefont {Kotov}}, \bibinfo {author} {\bibfnamefont {Stephen}\
  \bibnamefont {O'Brien}}, \ and\ \bibinfo {author} {\bibfnamefont
  {Christopher~B.}\ \bibnamefont {Murray}},\ }\bibfield  {title} {\enquote
  {\bibinfo {title} {Structural diversity in binary nanoparticle
  superlattices},}\ }\href {\doibase 10.1038/nature04414} {\bibfield  {journal}
  {\bibinfo  {journal} {Nature}\ }\textbf {\bibinfo {volume} {439}},\ \bibinfo
  {pages} {55--59} (\bibinfo {year} {2006})}\BibitemShut {NoStop}%
\bibitem [{\citenamefont {Hynninen}\ \emph {et~al.}(2007)\citenamefont
  {Hynninen}, \citenamefont {Thijssen}, \citenamefont {Vermolen}, \citenamefont
  {Dijkstra},\ and\ \citenamefont {{van Blaaderen}}}]{Hynninen:07}%
  \BibitemOpen
  \bibfield  {author} {\bibinfo {author} {\bibfnamefont {Antti-Pekka}\
  \bibnamefont {Hynninen}}, \bibinfo {author} {\bibfnamefont {Job H.~J.}\
  \bibnamefont {Thijssen}}, \bibinfo {author} {\bibfnamefont {Esther C.~M.}\
  \bibnamefont {Vermolen}}, \bibinfo {author} {\bibfnamefont {Marjolein}\
  \bibnamefont {Dijkstra}}, \ and\ \bibinfo {author} {\bibfnamefont {Alfons}\
  \bibnamefont {{van Blaaderen}}},\ }\bibfield  {title} {\enquote {\bibinfo
  {title} {Self-assembly route for photonic crystals with a bandgap in the
  visible region},}\ }\href {\doibase 10.1038/nmat1841} {\bibfield  {journal}
  {\bibinfo  {journal} {Nat. Mater.}\ }\textbf {\bibinfo {volume} {6}},\
  \bibinfo {pages} {202--205} (\bibinfo {year} {2007})}\BibitemShut {NoStop}%
\bibitem [{\citenamefont {Schaertl}\ \emph {et~al.}(2018)\citenamefont
  {Schaertl}, \citenamefont {Botin}, \citenamefont {Palberg},\ and\
  \citenamefont {Bartsch}}]{Schaertl:18}%
  \BibitemOpen
  \bibfield  {author} {\bibinfo {author} {\bibfnamefont {N.}~\bibnamefont
  {Schaertl}}, \bibinfo {author} {\bibfnamefont {D.}~\bibnamefont {Botin}},
  \bibinfo {author} {\bibfnamefont {T.}~\bibnamefont {Palberg}}, \ and\
  \bibinfo {author} {\bibfnamefont {E.}~\bibnamefont {Bartsch}},\ }\bibfield
  {title} {\enquote {\bibinfo {title} {Formation of {{Laves}} phases in
  buoyancy matched hard sphere suspensions},}\ }\href {\doibase
  10.1039/C7SM02348K} {\bibfield  {journal} {\bibinfo  {journal} {Soft Matter}\
  }\textbf {\bibinfo {volume} {14}},\ \bibinfo {pages} {5130--5139} (\bibinfo
  {year} {2018})}\BibitemShut {NoStop}%
\bibitem [{\citenamefont {Botet}\ \emph {et~al.}(2016)\citenamefont {Botet},
  \citenamefont {Cabane}, \citenamefont {Goehring}, \citenamefont {Li},\ and\
  \citenamefont {Artzner}}]{Botet:16}%
  \BibitemOpen
  \bibfield  {author} {\bibinfo {author} {\bibfnamefont {Robert}\ \bibnamefont
  {Botet}}, \bibinfo {author} {\bibfnamefont {Bernard}\ \bibnamefont {Cabane}},
  \bibinfo {author} {\bibfnamefont {Lucas}\ \bibnamefont {Goehring}}, \bibinfo
  {author} {\bibfnamefont {Joaquim}\ \bibnamefont {Li}}, \ and\ \bibinfo
  {author} {\bibfnamefont {Franck}\ \bibnamefont {Artzner}},\ }\bibfield
  {title} {\enquote {\bibinfo {title} {How do polydisperse repulsive colloids
  crystallize?}}\ }\href {\doibase 10.1039/C5FD00145E} {\bibfield  {journal}
  {\bibinfo  {journal} {Faraday Discuss.}\ }\textbf {\bibinfo {volume} {186}},\
  \bibinfo {pages} {229--240} (\bibinfo {year} {2016})}\BibitemShut {NoStop}%
\bibitem [{\citenamefont {Coslovich}\ \emph {et~al.}(2018)\citenamefont
  {Coslovich}, \citenamefont {Ozawa},\ and\ \citenamefont
  {Berthier}}]{Coslovich:18}%
  \BibitemOpen
  \bibfield  {author} {\bibinfo {author} {\bibfnamefont {Daniele}\ \bibnamefont
  {Coslovich}}, \bibinfo {author} {\bibfnamefont {Misaki}\ \bibnamefont
  {Ozawa}}, \ and\ \bibinfo {author} {\bibfnamefont {Ludovic}\ \bibnamefont
  {Berthier}},\ }\bibfield  {title} {\enquote {\bibinfo {title} {Local order
  and crystallization of dense polydisperse hard spheres},}\ }\href {\doibase
  10.1088/1361-648X/aab0c9} {\bibfield  {journal} {\bibinfo  {journal} {J.
  Phys.: Condens. Matter}\ }\textbf {\bibinfo {volume} {30}},\ \bibinfo {pages}
  {144004} (\bibinfo {year} {2018})}\BibitemShut {NoStop}%
\bibitem [{\citenamefont {Lindquist}\ \emph {et~al.}(2018)\citenamefont
  {Lindquist}, \citenamefont {Jadrich},\ and\ \citenamefont
  {Truskett}}]{Lindquist:18}%
  \BibitemOpen
  \bibfield  {author} {\bibinfo {author} {\bibfnamefont {Beth~A.}\ \bibnamefont
  {Lindquist}}, \bibinfo {author} {\bibfnamefont {Ryan~B.}\ \bibnamefont
  {Jadrich}}, \ and\ \bibinfo {author} {\bibfnamefont {Thomas~M.}\ \bibnamefont
  {Truskett}},\ }\bibfield  {title} {\enquote {\bibinfo {title} {Communication:
  {{From}} close-packed to topologically close-packed: {{Formation}} of
  {{Laves}} phases in moderately polydisperse hard-sphere mixtures},}\ }\href
  {\doibase 10.1063/1.5028279} {\bibfield  {journal} {\bibinfo  {journal} {J.
  Chem. Phys.}\ }\textbf {\bibinfo {volume} {148}},\ \bibinfo {pages} {191101}
  (\bibinfo {year} {2018})}\BibitemShut {NoStop}%
\bibitem [{\citenamefont {Fern{\'a}ndez}\ \emph {et~al.}(2007)\citenamefont
  {Fern{\'a}ndez}, \citenamefont {{Mart{\'i}n-Mayor}},\ and\ \citenamefont
  {Verrocchio}}]{Fernandez:07}%
  \BibitemOpen
  \bibfield  {author} {\bibinfo {author} {\bibfnamefont {L.~A.}\ \bibnamefont
  {Fern{\'a}ndez}}, \bibinfo {author} {\bibfnamefont {V.}~\bibnamefont
  {{Mart{\'i}n-Mayor}}}, \ and\ \bibinfo {author} {\bibfnamefont
  {P.}~\bibnamefont {Verrocchio}},\ }\bibfield  {title} {\enquote {\bibinfo
  {title} {Phase {{Diagram}} of a {{Polydisperse Soft}}-{{Spheres Model}} for
  {{Liquids}} and {{Colloids}}},}\ }\href {\doibase
  10.1103/PhysRevLett.98.085702} {\bibfield  {journal} {\bibinfo  {journal}
  {Phys Rev Lett}\ }\textbf {\bibinfo {volume} {98}},\ \bibinfo {pages}
  {085702} (\bibinfo {year} {2007})}\BibitemShut {NoStop}%
\bibitem [{\citenamefont {Byelov}\ \emph {et~al.}(2010)\citenamefont {Byelov},
  \citenamefont {Mourad}, \citenamefont {Snigireva}, \citenamefont {Snigirev},
  \citenamefont {Petukhov},\ and\ \citenamefont {Lekkerkerker}}]{Byelov:10}%
  \BibitemOpen
  \bibfield  {author} {\bibinfo {author} {\bibfnamefont {D.~V.}\ \bibnamefont
  {Byelov}}, \bibinfo {author} {\bibfnamefont {M.~C.~D.}\ \bibnamefont
  {Mourad}}, \bibinfo {author} {\bibfnamefont {I.}~\bibnamefont {Snigireva}},
  \bibinfo {author} {\bibfnamefont {A.}~\bibnamefont {Snigirev}}, \bibinfo
  {author} {\bibfnamefont {A.~V.}\ \bibnamefont {Petukhov}}, \ and\ \bibinfo
  {author} {\bibfnamefont {H.~N.~W.}\ \bibnamefont {Lekkerkerker}},\ }\bibfield
   {title} {\enquote {\bibinfo {title} {Experimental {{Observation}} of
  {{Fractionated Crystallization}} in {{Polydisperse Platelike Colloids}}},}\
  }\href {\doibase 10.1021/la100993k} {\bibfield  {journal} {\bibinfo
  {journal} {Langmuir}\ }\textbf {\bibinfo {volume} {26}},\ \bibinfo {pages}
  {6898--6901} (\bibinfo {year} {2010})}\BibitemShut {NoStop}%
\bibitem [{Not()}]{NoteSM}%
  \BibitemOpen
  \href@noop {} {}\bibinfo {note} {See Supplemental Material at http://XX,
  which includes Refs. [8,13,14,20,25,33--35,38--48], for an extended materials
  and methods section, as well as for details of the suspension size
  distribution, interdiffusion experiments, phase composition, glassy state,
  simulation convergence and lattice simulations.}\BibitemShut {Stop}%
\bibitem [{\citenamefont {Hynninen}\ and\ \citenamefont
  {Dijkstra}(2003)}]{Hynninen:03}%
  \BibitemOpen
  \bibfield  {author} {\bibinfo {author} {\bibfnamefont {Antti-Pekka}\
  \bibnamefont {Hynninen}}\ and\ \bibinfo {author} {\bibfnamefont {Marjolein}\
  \bibnamefont {Dijkstra}},\ }\bibfield  {title} {\enquote {\bibinfo {title}
  {Phase diagrams of hard-core repulsive {{Yukawa}} particles},}\ }\href
  {\doibase 10.1103/PhysRevE.68.021407} {\bibfield  {journal} {\bibinfo
  {journal} {Phys. Rev. E}\ }\textbf {\bibinfo {volume} {68}},\ \bibinfo
  {pages} {021407} (\bibinfo {year} {2003})}\BibitemShut {NoStop}%
\bibitem [{\citenamefont {Bareigts}\ and\ \citenamefont
  {Labbez}(2018)}]{Bareigts:18}%
  \BibitemOpen
  \bibfield  {author} {\bibinfo {author} {\bibfnamefont {Guillaume}\
  \bibnamefont {Bareigts}}\ and\ \bibinfo {author} {\bibfnamefont {Christophe}\
  \bibnamefont {Labbez}},\ }\bibfield  {title} {\enquote {\bibinfo {title}
  {Jellium and cell model for titratable colloids with continuous size
  distribution},}\ }\href {\doibase 10.1063/1.5066074} {\bibfield  {journal}
  {\bibinfo  {journal} {J. Chem. Phys.}\ }\textbf {\bibinfo {volume} {149}},\
  \bibinfo {pages} {244903} (\bibinfo {year} {2018})}\BibitemShut {NoStop}%
\bibitem [{\citenamefont {J{\"o}nsson}\ \emph {et~al.}(2011)\citenamefont
  {J{\"o}nsson}, \citenamefont {Persello}, \citenamefont {Li},\ and\
  \citenamefont {Cabane}}]{Jonsson:11a}%
  \BibitemOpen
  \bibfield  {author} {\bibinfo {author} {\bibfnamefont {Bo}~\bibnamefont
  {J{\"o}nsson}}, \bibinfo {author} {\bibfnamefont {J.}~\bibnamefont
  {Persello}}, \bibinfo {author} {\bibfnamefont {J.}~\bibnamefont {Li}}, \ and\
  \bibinfo {author} {\bibfnamefont {B.}~\bibnamefont {Cabane}},\ }\bibfield
  {title} {\enquote {\bibinfo {title} {Equation of {{State}} of {{Colloidal
  Dispersions}}},}\ }\href {\doibase 10.1021/la2001392} {\bibfield  {journal}
  {\bibinfo  {journal} {Langmuir}\ }\textbf {\bibinfo {volume} {27}},\ \bibinfo
  {pages} {6606--6614} (\bibinfo {year} {2011})}\BibitemShut {NoStop}%
\bibitem [{\citenamefont {Li}\ \emph {et~al.}(2015)\citenamefont {Li},
  \citenamefont {Turesson}, \citenamefont {Haglund}, \citenamefont {Cabane},\
  and\ \citenamefont {Skep{\"o}}}]{Li:15}%
  \BibitemOpen
  \bibfield  {author} {\bibinfo {author} {\bibfnamefont {Joaquim}\ \bibnamefont
  {Li}}, \bibinfo {author} {\bibfnamefont {Martin}\ \bibnamefont {Turesson}},
  \bibinfo {author} {\bibfnamefont {Caroline~Anderberg}\ \bibnamefont
  {Haglund}}, \bibinfo {author} {\bibfnamefont {Bernard}\ \bibnamefont
  {Cabane}}, \ and\ \bibinfo {author} {\bibfnamefont {Marie}\ \bibnamefont
  {Skep{\"o}}},\ }\bibfield  {title} {\enquote {\bibinfo {title} {Equation of
  state of {{PEG}}/{{PEO}} in good solvent. {{Comparison}} between a
  one-parameter {{EOS}} and experiments},}\ }\href {\doibase
  10.1016/j.polymer.2015.10.056} {\bibfield  {journal} {\bibinfo  {journal}
  {Polymer}\ }\textbf {\bibinfo {volume} {80}},\ \bibinfo {pages} {205--213}
  (\bibinfo {year} {2015})}\BibitemShut {NoStop}%
\bibitem [{\citenamefont {Goehring}\ \emph {et~al.}(2017)\citenamefont
  {Goehring}, \citenamefont {Li},\ and\ \citenamefont
  {Kiatkirakajorn}}]{Goehring:17}%
  \BibitemOpen
  \bibfield  {author} {\bibinfo {author} {\bibfnamefont {Lucas}\ \bibnamefont
  {Goehring}}, \bibinfo {author} {\bibfnamefont {Joaquim}\ \bibnamefont {Li}},
  \ and\ \bibinfo {author} {\bibfnamefont {Pree-Cha}\ \bibnamefont
  {Kiatkirakajorn}},\ }\bibfield  {title} {\enquote {\bibinfo {title} {Drying
  paint: From micro-scale dynamics to mechanical instabilities},}\ }\href
  {\doibase 10.1098/rsta.2016.0161} {\bibfield  {journal} {\bibinfo  {journal}
  {Phil. Trans. R. Soc. A}\ }\textbf {\bibinfo {volume} {375}},\ \bibinfo
  {pages} {20160161} (\bibinfo {year} {2017})}\BibitemShut {NoStop}%
\bibitem [{\citenamefont {Narayanan}\ \emph {et~al.}(2018)\citenamefont
  {Narayanan}, \citenamefont {Sztucki}, \citenamefont {Van~Vaerenbergh},
  \citenamefont {L{\ifmmode\acute{e}\else\'{e}\fi}onardon}, \citenamefont
  {Gorini}, \citenamefont {Claustre}, \citenamefont {Sever}, \citenamefont
  {Morse},\ and\ \citenamefont {Boesecke}}]{Narayanan:18}%
  \BibitemOpen
  \bibfield  {author} {\bibinfo {author} {\bibfnamefont {T.}~\bibnamefont
  {Narayanan}}, \bibinfo {author} {\bibfnamefont {M.}~\bibnamefont {Sztucki}},
  \bibinfo {author} {\bibfnamefont {P.}~\bibnamefont {Van~Vaerenbergh}},
  \bibinfo {author} {\bibfnamefont {J.}~\bibnamefont
  {L{\ifmmode\acute{e}\else\'{e}\fi}onardon}}, \bibinfo {author} {\bibfnamefont
  {J.}~\bibnamefont {Gorini}}, \bibinfo {author} {\bibfnamefont
  {L.}~\bibnamefont {Claustre}}, \bibinfo {author} {\bibfnamefont
  {F.}~\bibnamefont {Sever}}, \bibinfo {author} {\bibfnamefont
  {J.}~\bibnamefont {Morse}}, \ and\ \bibinfo {author} {\bibfnamefont
  {P.}~\bibnamefont {Boesecke}},\ }\bibfield  {title} {\enquote {\bibinfo
  {title} {{A multipurpose instrument for time-resolved ultra-small-angle and
  coherent X-ray scattering}},}\ }\href {\doibase 10.1107/S1600576718012748}
  {\bibfield  {journal} {\bibinfo  {journal} {J. Appl. Crystallogr.}\ }\textbf
  {\bibinfo {volume} {51}},\ \bibinfo {pages} {1511--1524} (\bibinfo {year}
  {2018})}\BibitemShut {NoStop}%
\bibitem [{\citenamefont {Goertz}\ \emph {et~al.}(2009)\citenamefont {Goertz},
  \citenamefont {Dingenouts},\ and\ \citenamefont {Nirschl}}]{Goertz:09}%
  \BibitemOpen
  \bibfield  {author} {\bibinfo {author} {\bibfnamefont {Verena}\ \bibnamefont
  {Goertz}}, \bibinfo {author} {\bibfnamefont {Nico}\ \bibnamefont
  {Dingenouts}}, \ and\ \bibinfo {author} {\bibfnamefont {Hermann}\
  \bibnamefont {Nirschl}},\ }\bibfield  {title} {\enquote {\bibinfo {title}
  {Comparison of {{Nanometric Particle Size Distributions}} as {{Determined}}
  by {{SAXS}}, {{TEM}} and {{Analytical Ultracentrifuge}}},}\ }\href {\doibase
  10.1002/ppsc.200800002} {\bibfield  {journal} {\bibinfo  {journal} {Part.
  Part. Syst. Charact.}\ }\textbf {\bibinfo {volume} {26}},\ \bibinfo {pages}
  {17--24} (\bibinfo {year} {2009})}\BibitemShut {NoStop}%
\bibitem [{\citenamefont {Shabalin}\ \emph {et~al.}(2016)\citenamefont
  {Shabalin}, \citenamefont {Meijer}, \citenamefont {Dronyak}, \citenamefont
  {Yefanov}, \citenamefont {Singer}, \citenamefont {Kurta}, \citenamefont
  {Lorenz}, \citenamefont {Gorobtsov}, \citenamefont {Dzhigaev}, \citenamefont
  {Kalbfleisch}, \citenamefont {Gulden}, \citenamefont {Zozulya}, \citenamefont
  {Sprung}, \citenamefont {Petukhov},\ and\ \citenamefont
  {Vartanyants}}]{Shabalin:16}%
  \BibitemOpen
  \bibfield  {author} {\bibinfo {author} {\bibfnamefont {A.~G.}\ \bibnamefont
  {Shabalin}}, \bibinfo {author} {\bibfnamefont {J.-M.}\ \bibnamefont
  {Meijer}}, \bibinfo {author} {\bibfnamefont {R.}~\bibnamefont {Dronyak}},
  \bibinfo {author} {\bibfnamefont {O.~M.}\ \bibnamefont {Yefanov}}, \bibinfo
  {author} {\bibfnamefont {A.}~\bibnamefont {Singer}}, \bibinfo {author}
  {\bibfnamefont {R.~P.}\ \bibnamefont {Kurta}}, \bibinfo {author}
  {\bibfnamefont {U.}~\bibnamefont {Lorenz}}, \bibinfo {author} {\bibfnamefont
  {O.~Y.}\ \bibnamefont {Gorobtsov}}, \bibinfo {author} {\bibfnamefont
  {D.}~\bibnamefont {Dzhigaev}}, \bibinfo {author} {\bibfnamefont
  {S.}~\bibnamefont {Kalbfleisch}}, \bibinfo {author} {\bibfnamefont
  {J.}~\bibnamefont {Gulden}}, \bibinfo {author} {\bibfnamefont {A.~V.}\
  \bibnamefont {Zozulya}}, \bibinfo {author} {\bibfnamefont {M.}~\bibnamefont
  {Sprung}}, \bibinfo {author} {\bibfnamefont {A.~V.}\ \bibnamefont
  {Petukhov}}, \ and\ \bibinfo {author} {\bibfnamefont {I.~A.}\ \bibnamefont
  {Vartanyants}},\ }\bibfield  {title} {\enquote {\bibinfo {title} {Revealing
  {{Three}}-{{Dimensional Structure}} of an {{Individual Colloidal Crystal
  Grain}} by {{Coherent X}}-{{Ray Diffractive Imaging}}},}\ }\href {\doibase
  10.1103/PhysRevLett.117.138002} {\bibfield  {journal} {\bibinfo  {journal}
  {Phys. Rev. Lett.}\ }\textbf {\bibinfo {volume} {117}},\ \bibinfo {pages}
  {138002} (\bibinfo {year} {2016})}\BibitemShut {NoStop}%
\bibitem [{\citenamefont {Monovoukas}\ and\ \citenamefont
  {Gast}(1989)}]{Monovoukas:89}%
  \BibitemOpen
  \bibfield  {author} {\bibinfo {author} {\bibfnamefont {Yiannis}\ \bibnamefont
  {Monovoukas}}\ and\ \bibinfo {author} {\bibfnamefont {Alice~P}\ \bibnamefont
  {Gast}},\ }\bibfield  {title} {\enquote {\bibinfo {title} {The experimental
  phase diagram of charged colloidal suspensions},}\ }\href {\doibase
  10.1016/0021-9797(89)90368-8} {\bibfield  {journal} {\bibinfo  {journal}
  {Journal of Colloid and Interface Science}\ }\textbf {\bibinfo {volume}
  {128}},\ \bibinfo {pages} {533--548} (\bibinfo {year} {1989})}\BibitemShut
  {NoStop}%
\bibitem [{\citenamefont {Sirota}\ \emph {et~al.}(1989)\citenamefont {Sirota},
  \citenamefont {{Ou-Yang}}, \citenamefont {Sinha}, \citenamefont {Chaikin},
  \citenamefont {Axe},\ and\ \citenamefont {Fujii}}]{Sirota:89}%
  \BibitemOpen
  \bibfield  {author} {\bibinfo {author} {\bibfnamefont {E.~B.}\ \bibnamefont
  {Sirota}}, \bibinfo {author} {\bibfnamefont {H.~D.}\ \bibnamefont
  {{Ou-Yang}}}, \bibinfo {author} {\bibfnamefont {S.~K.}\ \bibnamefont
  {Sinha}}, \bibinfo {author} {\bibfnamefont {P.~M.}\ \bibnamefont {Chaikin}},
  \bibinfo {author} {\bibfnamefont {J.~D.}\ \bibnamefont {Axe}}, \ and\
  \bibinfo {author} {\bibfnamefont {Y.}~\bibnamefont {Fujii}},\ }\bibfield
  {title} {\enquote {\bibinfo {title} {Complete phase diagram of a charged
  colloidal system: {{A}} synchrotron x-ray scattering study},}\ }\href
  {\doibase 10.1103/PhysRevLett.62.1524} {\bibfield  {journal} {\bibinfo
  {journal} {Phys. Rev. Lett.}\ }\textbf {\bibinfo {volume} {62}},\ \bibinfo
  {pages} {1524--1527} (\bibinfo {year} {1989})}\BibitemShut {NoStop}%
\bibitem [{\citenamefont {Versmold}(1995)}]{Versmold:95}%
  \BibitemOpen
  \bibfield  {author} {\bibinfo {author} {\bibfnamefont {Heiner}\ \bibnamefont
  {Versmold}},\ }\bibfield  {title} {\enquote {\bibinfo {title} {Neutron
  {{Diffraction}} from {{Shear Ordered Colloidal Dispersions}}},}\ }\href
  {\doibase 10.1103/PhysRevLett.75.763} {\bibfield  {journal} {\bibinfo
  {journal} {Phys. Rev. Lett.}\ }\textbf {\bibinfo {volume} {75}},\ \bibinfo
  {pages} {763--766} (\bibinfo {year} {1995})}\BibitemShut {NoStop}%
\bibitem [{\citenamefont {Gottwald}\ \emph {et~al.}(2004)\citenamefont
  {Gottwald}, \citenamefont {Likos}, \citenamefont {Kahl},\ and\ \citenamefont
  {L{\"o}wen}}]{Gottwald:04}%
  \BibitemOpen
  \bibfield  {author} {\bibinfo {author} {\bibfnamefont {D.}~\bibnamefont
  {Gottwald}}, \bibinfo {author} {\bibfnamefont {C.~N.}\ \bibnamefont {Likos}},
  \bibinfo {author} {\bibfnamefont {G.}~\bibnamefont {Kahl}}, \ and\ \bibinfo
  {author} {\bibfnamefont {H.}~\bibnamefont {L{\"o}wen}},\ }\bibfield  {title}
  {\enquote {\bibinfo {title} {Phase {{Behavior}} of {{Ionic Microgels}}},}\
  }\href {\doibase 10.1103/PhysRevLett.92.068301} {\bibfield  {journal}
  {\bibinfo  {journal} {Phys. Rev. Lett.}\ }\textbf {\bibinfo {volume} {92}},\
  \bibinfo {pages} {068301} (\bibinfo {year} {2004})}\BibitemShut {NoStop}%
\bibitem [{\citenamefont {P{\`a}mies}\ \emph {et~al.}(2009)\citenamefont
  {P{\`a}mies}, \citenamefont {Cacciuto},\ and\ \citenamefont
  {Frenkel}}]{Pamies:09}%
  \BibitemOpen
  \bibfield  {author} {\bibinfo {author} {\bibfnamefont {Josep~C.}\
  \bibnamefont {P{\`a}mies}}, \bibinfo {author} {\bibfnamefont {Angelo}\
  \bibnamefont {Cacciuto}}, \ and\ \bibinfo {author} {\bibfnamefont {Daan}\
  \bibnamefont {Frenkel}},\ }\bibfield  {title} {\enquote {\bibinfo {title}
  {Phase diagram of {{Hertzian}} spheres},}\ }\href {\doibase
  10.1063/1.3186742} {\bibfield  {journal} {\bibinfo  {journal} {J. Chem.
  Phys.}\ }\textbf {\bibinfo {volume} {131}},\ \bibinfo {pages} {044514}
  (\bibinfo {year} {2009})}\BibitemShut {NoStop}%
\bibitem [{\citenamefont {Mohanty}\ and\ \citenamefont
  {Richtering}(2008)}]{Mohanty:08}%
  \BibitemOpen
  \bibfield  {author} {\bibinfo {author} {\bibfnamefont {P.~S.}\ \bibnamefont
  {Mohanty}}\ and\ \bibinfo {author} {\bibfnamefont {W.}~\bibnamefont
  {Richtering}},\ }\bibfield  {title} {\enquote {\bibinfo {title} {Structural
  {{Ordering}} and {{Phase Behavior}} of {{Charged Microgels}}},}\ }\href
  {\doibase 10.1021/jp808203d} {\bibfield  {journal} {\bibinfo  {journal} {J.
  Phys. Chem. B}\ }\textbf {\bibinfo {volume} {112}},\ \bibinfo {pages}
  {14692--14697} (\bibinfo {year} {2008})}\BibitemShut {NoStop}%
\bibitem [{\citenamefont {Grigera}\ and\ \citenamefont
  {Parisi}(2001)}]{Grigera:01}%
  \BibitemOpen
  \bibfield  {author} {\bibinfo {author} {\bibfnamefont {Tom{\'a}s~S.}\
  \bibnamefont {Grigera}}\ and\ \bibinfo {author} {\bibfnamefont {Giorgio}\
  \bibnamefont {Parisi}},\ }\bibfield  {title} {\enquote {\bibinfo {title}
  {Fast {{Monte Carlo}} algorithm for supercooled soft spheres},}\ }\href
  {\doibase 10.1103/PhysRevE.63.045102} {\bibfield  {journal} {\bibinfo
  {journal} {Phys Rev E}\ }\textbf {\bibinfo {volume} {63}},\ \bibinfo {pages}
  {045102(R)} (\bibinfo {year} {2001})}\BibitemShut {NoStop}%
\bibitem [{\citenamefont {Brito}\ \emph {et~al.}(2018)\citenamefont {Brito},
  \citenamefont {Lerner},\ and\ \citenamefont {Wyart}}]{Brito:18}%
  \BibitemOpen
  \bibfield  {author} {\bibinfo {author} {\bibfnamefont {Carolina}\
  \bibnamefont {Brito}}, \bibinfo {author} {\bibfnamefont {Edan}\ \bibnamefont
  {Lerner}}, \ and\ \bibinfo {author} {\bibfnamefont {Matthieu}\ \bibnamefont
  {Wyart}},\ }\bibfield  {title} {\enquote {\bibinfo {title} {Theory for {{Swap
  Acceleration}} near the {{Glass}} and {{Jamming Transitions}} for
  {{Continuously Polydisperse Particles}}},}\ }\href {\doibase
  10.1103/PhysRevX.8.031050} {\bibfield  {journal} {\bibinfo  {journal} {Phys.
  Rev. X}\ }\textbf {\bibinfo {volume} {8}},\ \bibinfo {pages} {031050}
  (\bibinfo {year} {2018})}\BibitemShut {NoStop}%
\bibitem [{\citenamefont {Lechner}\ and\ \citenamefont
  {Dellago}(2008)}]{Lechner:08}%
  \BibitemOpen
  \bibfield  {author} {\bibinfo {author} {\bibfnamefont {Wolfgang}\
  \bibnamefont {Lechner}}\ and\ \bibinfo {author} {\bibfnamefont {Christoph}\
  \bibnamefont {Dellago}},\ }\bibfield  {title} {\enquote {\bibinfo {title}
  {Accurate determination of crystal structures based on averaged local bond
  order parameters},}\ }\href {\doibase 10.1063/1.2977970} {\bibfield
  {journal} {\bibinfo  {journal} {The Journal of Chemical Physics}\ }\textbf
  {\bibinfo {volume} {129}},\ \bibinfo {pages} {114707} (\bibinfo {year}
  {2008})}\BibitemShut {NoStop}%
\bibitem [{\citenamefont {Kiatkirakajorn}(2018)}]{Kiatkirakajorn:18}%
  \BibitemOpen
  \bibfield  {author} {\bibinfo {author} {\bibfnamefont {Pree-Cha}\
  \bibnamefont {Kiatkirakajorn}},\ }\emph {\bibinfo {title} {Morphological
  Instabilities in Drying Colloids}},\ \href@noop {} {\bibinfo {type} {{{PhD
  Thesis}}}},\ \bibinfo  {school} {Georg-August University}, \bibinfo {address}
  {{G{\"o}ttingen}} (\bibinfo {year} {2018})\BibitemShut {NoStop}%
\bibitem [{\citenamefont {Hamaguchi}\ \emph {et~al.}(1997)\citenamefont
  {Hamaguchi}, \citenamefont {Farouki},\ and\ \citenamefont
  {Dubin}}]{Hamaguchi:97}%
  \BibitemOpen
  \bibfield  {author} {\bibinfo {author} {\bibfnamefont {S.}~\bibnamefont
  {Hamaguchi}}, \bibinfo {author} {\bibfnamefont {R.~T.}\ \bibnamefont
  {Farouki}}, \ and\ \bibinfo {author} {\bibfnamefont {D.~H.~E.}\ \bibnamefont
  {Dubin}},\ }\bibfield  {title} {\enquote {\bibinfo {title} {Triple point of
  {{Yukawa}} systems},}\ }\href {\doibase 10.1103/PhysRevE.56.4671} {\bibfield
  {journal} {\bibinfo  {journal} {Phys. Rev. E}\ }\textbf {\bibinfo {volume}
  {56}},\ \bibinfo {pages} {4671--4682} (\bibinfo {year} {1997})}\BibitemShut
  {NoStop}%
\bibitem [{\citenamefont {Botet}\ and\ \citenamefont
  {Cabane}(2012)}]{Botet:12a}%
  \BibitemOpen
  \bibfield  {author} {\bibinfo {author} {\bibfnamefont {Robert}\ \bibnamefont
  {Botet}}\ and\ \bibinfo {author} {\bibfnamefont {Bernard}\ \bibnamefont
  {Cabane}},\ }\bibfield  {title} {\enquote {\bibinfo {title} {Simple inversion
  formula for the small-angle {{X}}-ray scattering intensity from polydisperse
  systems of spheres},}\ }\href {\doibase 10.1107/S0021889812012812} {\bibfield
   {journal} {\bibinfo  {journal} {J. Appl. Crystallogr.}\ }\textbf {\bibinfo
  {volume} {45}},\ \bibinfo {pages} {406--416} (\bibinfo {year}
  {2012})}\BibitemShut {NoStop}%
\bibitem [{\citenamefont {Li}\ \emph {et~al.}(2012)\citenamefont {Li},
  \citenamefont {Cabane}, \citenamefont {Sztucki}, \citenamefont {Gummel},\
  and\ \citenamefont {Goehring}}]{Li:12}%
  \BibitemOpen
  \bibfield  {author} {\bibinfo {author} {\bibfnamefont {Joaquim}\ \bibnamefont
  {Li}}, \bibinfo {author} {\bibfnamefont {Bernard}\ \bibnamefont {Cabane}},
  \bibinfo {author} {\bibfnamefont {Michael}\ \bibnamefont {Sztucki}}, \bibinfo
  {author} {\bibfnamefont {J{\'e}r{\'e}mie}\ \bibnamefont {Gummel}}, \ and\
  \bibinfo {author} {\bibfnamefont {Lucas}\ \bibnamefont {Goehring}},\
  }\bibfield  {title} {\enquote {\bibinfo {title} {Drying {{Dip}}-{{Coated
  Colloidal Films}}},}\ }\href {\doibase 10.1021/la203549g} {\bibfield
  {journal} {\bibinfo  {journal} {Langmuir}\ }\textbf {\bibinfo {volume}
  {28}},\ \bibinfo {pages} {200--208} (\bibinfo {year} {2012})}\BibitemShut
  {NoStop}%
\bibitem [{\citenamefont {Boon}\ \emph {et~al.}(2015)\citenamefont {Boon},
  \citenamefont {{Guerrero-Garc{\'i}a}}, \citenamefont {van Roij},\ and\
  \citenamefont {de~la Cruz}}]{Boon:15}%
  \BibitemOpen
  \bibfield  {author} {\bibinfo {author} {\bibfnamefont {Niels}\ \bibnamefont
  {Boon}}, \bibinfo {author} {\bibfnamefont {Guillermo~Ivan}\ \bibnamefont
  {{Guerrero-Garc{\'i}a}}}, \bibinfo {author} {\bibfnamefont {Ren{\'e}}\
  \bibnamefont {van Roij}}, \ and\ \bibinfo {author} {\bibfnamefont
  {Monica~Olvera}\ \bibnamefont {de~la Cruz}},\ }\bibfield  {title} {\enquote
  {\bibinfo {title} {Effective charges and virial pressure of concentrated
  macroion solutions},}\ }\href {\doibase 10.1073/pnas.1511798112} {\bibfield
  {journal} {\bibinfo  {journal} {PNAS}\ }\textbf {\bibinfo {volume} {112}},\
  \bibinfo {pages} {9242--9246} (\bibinfo {year} {2015})}\BibitemShut {NoStop}%
\bibitem [{\citenamefont {Bolt}(1957)}]{Bolt:57}%
  \BibitemOpen
  \bibfield  {author} {\bibinfo {author} {\bibfnamefont {G.~H.}\ \bibnamefont
  {Bolt}},\ }\bibfield  {title} {\enquote {\bibinfo {title} {Determination of
  the {{Charge Density}} of {{Silica Sols}}},}\ }\href {\doibase
  10.1021/j150555a007} {\bibfield  {journal} {\bibinfo  {journal} {J. Phys.
  Chem.}\ }\textbf {\bibinfo {volume} {61}},\ \bibinfo {pages} {1166--1169}
  (\bibinfo {year} {1957})}\BibitemShut {NoStop}%
\bibitem [{\citenamefont {Labbez}\ \emph {et~al.}(2009)\citenamefont {Labbez},
  \citenamefont {Jonsson}, \citenamefont {Skarba},\ and\ \citenamefont
  {Borkovec}}]{Labbez:09}%
  \BibitemOpen
  \bibfield  {author} {\bibinfo {author} {\bibfnamefont {Christophe}\
  \bibnamefont {Labbez}}, \bibinfo {author} {\bibfnamefont {Bo}~\bibnamefont
  {Jonsson}}, \bibinfo {author} {\bibfnamefont {Michal}\ \bibnamefont
  {Skarba}}, \ and\ \bibinfo {author} {\bibfnamefont {Michal}\ \bibnamefont
  {Borkovec}},\ }\bibfield  {title} {\enquote {\bibinfo {title} {Ion- ion
  correlation and charge reversal at titrating solid interfaces},}\ }\href@noop
  {} {\bibfield  {journal} {\bibinfo  {journal} {Langmuir}\ }\textbf {\bibinfo
  {volume} {25}},\ \bibinfo {pages} {7209--7213} (\bibinfo {year}
  {2009})}\BibitemShut {NoStop}%
\bibitem [{\citenamefont {Dove}\ and\ \citenamefont {Craven}(2005)}]{Dove:05}%
  \BibitemOpen
  \bibfield  {author} {\bibinfo {author} {\bibfnamefont {P.~M.}\ \bibnamefont
  {Dove}}\ and\ \bibinfo {author} {\bibfnamefont {C.~M.}\ \bibnamefont
  {Craven}},\ }\bibfield  {title} {\enquote {\bibinfo {title} {Surface charge
  density on silica in alkali and alkaline earth chloride electrolyte
  solutions},}\ }\href@noop {} {\bibfield  {journal} {\bibinfo  {journal}
  {Geochim Cosmochim Acta}\ }\textbf {\bibinfo {volume} {69}},\ \bibinfo
  {pages} {4963--4970} (\bibinfo {year} {2005})}\BibitemShut {NoStop}%
\bibitem [{\citenamefont {Frenkel}\ and\ \citenamefont
  {Smit}(2002)}]{Frenkel:02}%
  \BibitemOpen
  \bibfield  {author} {\bibinfo {author} {\bibfnamefont {D.}~\bibnamefont
  {Frenkel}}\ and\ \bibinfo {author} {\bibfnamefont {B.}~\bibnamefont {Smit}},\
  }\href@noop {} {\emph {\bibinfo {title} {Understanding {{Molecular
  Simulation}}}}}\ (\bibinfo  {publisher} {{Academic Press}},\ \bibinfo
  {address} {San Diego},\ \bibinfo {year} {2002})\BibitemShut {NoStop}%
\bibitem [{\citenamefont {Wilding}\ and\ \citenamefont
  {Sollich}(2010)}]{Wilding:10}%
  \BibitemOpen
  \bibfield  {author} {\bibinfo {author} {\bibfnamefont {Nigel~B.}\
  \bibnamefont {Wilding}}\ and\ \bibinfo {author} {\bibfnamefont {Peter}\
  \bibnamefont {Sollich}},\ }\bibfield  {title} {\enquote {\bibinfo {title}
  {Phase behavior of polydisperse spheres: {{Simulation}} strategies and an
  application to the freezing transition},}\ }\href {\doibase
  10.1063/1.3510534} {\bibfield  {journal} {\bibinfo  {journal} {The Journal of
  Chemical Physics}\ }\textbf {\bibinfo {volume} {133}},\ \bibinfo {pages}
  {224102} (\bibinfo {year} {2010})}\BibitemShut {NoStop}%
\bibitem [{\citenamefont {Verlet}(1968)}]{Verlet:68}%
  \BibitemOpen
  \bibfield  {author} {\bibinfo {author} {\bibfnamefont {Loup}\ \bibnamefont
  {Verlet}},\ }\bibfield  {title} {\enquote {\bibinfo {title} {Computer
  "{{Experiments}}" on {{Classical Fluids}}. {{II}}. {{Equilibrium Correlation
  Functions}}},}\ }\href {\doibase 10.1103/PhysRev.165.201} {\bibfield
  {journal} {\bibinfo  {journal} {Phys. Rev.}\ }\textbf {\bibinfo {volume}
  {165}},\ \bibinfo {pages} {201--214} (\bibinfo {year} {1968})}\BibitemShut
  {NoStop}%
\bibitem [{\citenamefont {Hansen}\ and\ \citenamefont
  {Verlet}(1969)}]{Hansen:69}%
  \BibitemOpen
  \bibfield  {author} {\bibinfo {author} {\bibfnamefont {Jean-Pierre}\
  \bibnamefont {Hansen}}\ and\ \bibinfo {author} {\bibfnamefont {Loup}\
  \bibnamefont {Verlet}},\ }\bibfield  {title} {\enquote {\bibinfo {title}
  {Phase {{Transitions}} of the {{Lennard}}-{{Jones System}}},}\ }\href
  {\doibase 10.1103/PhysRev.184.151} {\bibfield  {journal} {\bibinfo  {journal}
  {Phys. Rev.}\ }\textbf {\bibinfo {volume} {184}},\ \bibinfo {pages}
  {151--161} (\bibinfo {year} {1969})}\BibitemShut {NoStop}%
\bibitem [{\citenamefont {Leocmach}\ \emph {et~al.}(2013)\citenamefont
  {Leocmach}, \citenamefont {Russo},\ and\ \citenamefont
  {Tanaka}}]{Leocmach:13}%
  \BibitemOpen
  \bibfield  {author} {\bibinfo {author} {\bibfnamefont {Mathieu}\ \bibnamefont
  {Leocmach}}, \bibinfo {author} {\bibfnamefont {John}\ \bibnamefont {Russo}},
  \ and\ \bibinfo {author} {\bibfnamefont {Hajime}\ \bibnamefont {Tanaka}},\
  }\bibfield  {title} {\enquote {\bibinfo {title} {Importance of many-body
  correlations in glass transition: {{An}} example from polydisperse hard
  spheres},}\ }\href {\doibase 10.1063/1.4769981} {\bibfield  {journal}
  {\bibinfo  {journal} {The Journal of Chemical Physics}\ }\textbf {\bibinfo
  {volume} {138}},\ \bibinfo {pages} {12A536} (\bibinfo {year}
  {2013})}\BibitemShut {NoStop}%
\end{thebibliography}%

\end{document}